\documentclass{article}[a4]
\usepackage[fleqn]{amsmath}
\usepackage{mathtools}

\renewcommand{\a}{\alpha} 
\renewcommand{\b}{\beta}
\newcommand{\g}{\gamma} 

\renewcommand{\d}{\delta}

\newcommand{\s}{\sigma}

\renewcommand{\t}{\tau}

\newcommand{\cO}{{\cal O}}

\newcommand{\half}{\frac{1}{2}}

\begin{document}

\author{Jiri Hoogland\footnote{jiri@cwi.nl, jiri.hoogland@gmail.com} 
  and Dimitri Neumann\footnote{neumann@cwi.nl}\\
  CWI, P.O.~Box 94079, 1090 GB  Amsterdam, The Netherlands}
\title{\textbf{Tradable Schemes}}
\maketitle

\thispagestyle{empty}
\begin{abstract}
In this article we present a new approach to the numerical
valuation of derivative securities. The method is based on
our previous work where we formulated the theory of pricing 
in terms of tradables. The basic idea is to fit a finite
difference scheme to exact solutions of the pricing PDE.
This can be done in a very elegant way, due to the fact that
in our tradable based formulation there appear no drift terms
in the PDE. We construct a mixed scheme based on this idea and
apply it to price various types of arithmetic Asian options, as
well as plain vanilla options (both european and american style)
on stocks paying known cash dividends. We find prices which are
accurate to $\sim 0.1\%$ in about 10ms on a Pentium 233MHz
computer and to $\sim 0.001\%$ in a second. The scheme can
also be used for market conform pricing, by fitting it to
observed option prices.
\end{abstract}

\newpage

\section{Introduction}

One of the most popular methods for pricing (exotic)
derivative securities is to make use of finite difference
schemes to solve the PDE associated with the pricing problem.
In a world where prices are driven by Wiener processes such PDE's
are usually of generalized diffusion type
\[ 
    \bigg( 
        \partial_t 
        + \a(x,t)
        + \b(x,t) \partial_x
        + \g(x,t) \partial_x^2 
    \bigg) 
    V(x,t) = 0 
\]
In such an equation the terms containing $\a,\b$ and $\g$
will be called the constant, drift and diffusion term respectively.
In the traditional formulation of option pricing theory,
the PDE to be solved generally contains not only diffusion terms
but also non-trivial constant and drift terms. The most obvious
example is of course the Black-Scholes PDE:
\[ 
    \bigg( 
        \partial_t 
        - r 
        + rS\partial_S 
        + \frac{1}{2}\sigma^2S^2\partial_S^2 
    \bigg) 
    V(S,t) = 0 
\]
It is well known that exactly the presence of drift terms in
the PDE can lead to severe numerical problems, unless they are
handled with much care.

In this article we present a new approach to the numerical
valuation of derivative securities. The approach is based on our
previous work \cite{HooglandNeumann00a,HooglandNeumann00b,
HooglandNeumann00c} where we introduced a new formalism for
contingent claim pricing. The core of this formalism is the idea
that problems should be formulated in terms of {\it tradable objects
only}. (Note that we use a broad definition of the term tradable:
every quantity that can be represented by a self-financing portfolio
is considered to be a tradable). From a numerical point of view,
the formalism has the distinct advantage that pricing PDE's never
contain drift nor constant terms. This makes it a very natural
starting point for the construction of finite difference schemes
for pricing derivative securities.

Our key idea is to require that a given set of solutions of the
underlying PDE should also be an exact solution of the numerical
scheme. We consider the case where there are two underlying tradables
in the pricing problem, leading to a one dimensional PDE. It then
turns out that we only need three exact solutions to fix a fully
implicit (or explicit) scheme.  By combining the fully implicit and
fully explicit schemes into a Crank-Nicolson type scheme, we obtain a
very powerful mixed scheme, still fitted to the exact solutions.
Details on convergence and stability properties will be treated in
future work.

So how do we fix the scheme? A very natural choice is to use the two
tradables themselves as exact solutions of the PDE. So we need only
one more non-trivial solution of the PDE. Such a solution is in
general not too hard to obtain, as we show in this article. Note that
instead of using exact solutions we could also use observed market
prices to fix the scheme, generalizing in a natural way the concept of
implied trees of Ref.~\cite{DermanKani94}. The latter idea will be
explored in future work.

In this article we present results on the performance of our scheme
when applied to the pricing of a well known problematic case: the
arithmetic Asian option, and furthermore results on the pricing of
plain vanilla options on stocks paying cash dividends, which are
related to arithmetic Asians by a symmetry operation (see
Ref.~\cite{HooglandNeumann00c}).  Note however that these applications
only serve as an example of the power of the our scheme. The scheme
itself can be used in a much broader class of pricing problems.

Let us recall some other approaches to the problem of pricing 
arithmetic Asians which can be found in the literature. To start with,
there are the (semi-)analytical approaches. The first result in this
direction was given in Ref.~\cite{GemanYor93}, who derived a closed
form expression for the Laplace-transform of the price of a
continuously sampled average price call. In our
paper~\cite{HooglandNeumann00c} we provided a similar solution
for the price of an arithmetic average strike put. Exploiting the
various symmetries underlying the pricing of contingent claims,
we furthermore showed that unseasoned arithmetic average price
and strike options are closely related objects, which can be
transformed into each other by a suitable substitution of parameters.
However, to find actual prices, one needs to calculate an inverse
Laplace transform. This turns out to be a hard problem, which
requires a substantial amount of numerical work~\cite{FuMadanWang98}
and makes the method less suitable for practical purposes.

For this reason, people usually revert to approximative or numerical
methods instead. These methods can be roughly classified in three
categories. The first category consists of analytical approximations.
In Ref.~\cite{TurnbullWakeman91} the first two moments of the
distribution of the arithmetic average are fitted to a lognormal
distribution. This leads to a Black-Scholes type pricing
formula. Alternatively, in Refs.~\cite{Curran94,RogersShi95,
Thompson99} quite accurate upper- and lower bounds on prices
are derived. These methods can be very fast. Disadvantages are
that they have a fixed accuracy and they can not easily be adapted
to handle discretely sampled Asians and early exercise features. 

The second category consists of methods which use the path integral or
Feynman-Kac formulation of the problem as a starting point. A path
integral can easily be evaluated using a Monte Carlo
approach~\cite{KemnaVorst92}. Just generate a large number of paths
and take averages over the payoffs associated with every individual
path. This method is versatile, simple to use and easy to implement.
Main disadvantages are that convergence is slow, and the incorporation
of early exercise features is notably hard. To speed up the
calculation, one can make use of a proper control-variate to reduce
the variance. In the case of arithmetic Asians, the obvious choice of
a control variate is a geometric Asian for which closed form formulas
do exist. Another disadvantage of Monte Carlo simulation is that paths
are sampled discretely. Therefore the method inherently makes a
discretization error when calculating continuously sampled Asians.

The third category consists of methods which numerically solve the PDE
associated to the pricing problem. This approach is flexible and early
exercise features can be incorporated in a straightforward manner. For
arithmetic Asians there exist various forms of this PDE (e.g.
Refs.~\cite{Ingersoll87,RogersShi95,Zhang01,HooglandNeumann00b}). In
principle all these forms can be related to each other by suitable
changes of variables. However, to the best of our knowledge, our
formulation of pricing theory in terms of tradables is the first and
only one which leads in a natural and straightforward way to a PDE
without drift and constant terms. Other approaches lead to PDE's which
are plagued by the presence of drift terms. The main problem
associated to the appearance of a drift term is that for certain
parameter ranges it tends to dominate the diffusion term. The PDE then
becomes of hyperbolic type instead of parabolic type.  If such a
problem is attacked using standard techniques for solving parabolic
(diffusion) equations, one finds spurious oscillations in the
numerical solution.  Similar problems are well known in the field of
computational fluid dynamics, where they are known as boundary layer
problems. It is this field in which methods were developed to
circumvent the problem. One approach is to make use of so-called
flux-delimiters. Their use in the pricing of arithmetic Asians was
introduced in Ref.~\cite{ZvanForsythVetzal97}. Flux-limiters may have 
their use in CFD, but they seem to be too much overkill for the problem 
at hand. By writing the pricing problem in the proper coordinates, 
i.e. tradables, the boundary layer problems can be trivially avoided 
since drift terms are absent by construction.

Using a finite-difference scheme will lead to an discretization-error. 
Our scheme takes care of this fact by fitting to an exact analytical solution.
The scheme is in effect what is known as a fitted scheme~\cite{Duffy,Morton96}. 
Fitted schemes have been used in the past to deal with boundary layer problems. 
Here the idea is to construct a finite difference scheme in such a way that stationary
(time-independent) solutions of the underlying PDE are exact solutions of the numerical 
scheme.

By fitting our FD scheme to an analytical solution of the PDE we can better correct 
for the discretization errors that will occur. For example when we compute the price of a discretely 
sampled arithmetic asian option the volatility function will make jumps at the sample times. 
In a standard FD scheme this will clearly lead to complications and errors.
In the fitted scheme,the scheme is exactly adapted to this jump because we use a tradable
that is a solution too.

The outline of this article is as follows. Section 2 gives some
background and motivation. In section 3 the construction of our
finite difference scheme is described. In section 4 we apply the
scheme to the PDE for Asian type options as well as vanilla options
on stock paying cash dividend. Section 5 provides numerical results
for both type of options. In section 6 we conclude and give an outlook. 

\section{Motivation}

In previous work \cite{HooglandNeumann00a,HooglandNeumann00b,
HooglandNeumann00c}, we have shown that in a market
with two tradables, say $S$ and $P$, where price processes are driven 
by Brownian motions, the pricing PDE can be reduced to the form
\[ 
    \bigg(
        -\partial_{\t}+\half\s(x,t)^2x^2\partial_x^2
    \bigg)
    V(x,\t) = 0 
\]
where $P$ was chosen as numeraire, $\t$ is time to maturity, $x:=
S/P$ and the price is $PV(x,\t)$.  Note the absence of a drift term.
Due to this fact, the PDE has the trivial solutions $x$ and $1$,
corresponding to the basic tradables $S$ and $P$ respectively. So
the pricing problem is fixed by knowledge of $\s(x,\t)$. Alternatively,
suppose we know one more solution of this equation, one that has a
non-vanishing second derivative with respect to $x$ (gamma), call it
$R(x,\t)$. Then we can write
\[ 
    \frac{1}{2}\s(x,\t)^2 x^2 = \frac{\partial_{\t}R(x,\t)}{\partial_x^2 R(x,\t)} 
\]
In other words, the knowledge of the price of one sufficiently
well-behaved derivative security also fixes the model. Let us consider
as an example the Black-Scholes context. It is easy to construct
a tradable with payoff $x^2$ at time $\t=0$ (We will call it a
quadratic tradable. It is a special case of a power tradable.
See Ref.~\cite{HooglandNeumann00b}). It is given by
\[ 
    x^2 e^{\s^2\t} 
\]
Plugging this into the formula we find
\[ 
    \s(x,t)=\s 
\]
as expected. Note that if we use observed option prices
of, say, call options, taking derivatives w.r.t. expiration
time and strike, we recover the concept of local volatility,
the basic idea behind implied trees \cite{DermanKani94}.

\section{Finite differences}

We can use the same line of thought to construct very efficient
finite difference schemes. The idea is to construct a scheme in
such a way that it has the underlying tradables as well as one
given derived tradable as exact solutions. We will now show that
these conditions uniquely fix a fully implicit scheme. Indeed,
a fully implicit scheme is defined by the relations
\[ 
    a \d_x^+ \d_x^- V(x,\t) 
    + b \d_x^0 V(x,\t)
    + c V(x,\t) 
    = d \d_\t^- V(x,\t) 
\]
where
\begin{align*} 
    \d_x^+ \d_x^- V(x,\t) &:= \frac{V(x+\d x,\t)-2V(x,\t)+V(x-\d x,\t)}{\d x^2} \\
    \d_x^0 V(x,\t) &:= \frac{V(x+\d x,\t)-V(x-\d x,\t)}{2\d x} \\
    \d_\t^- V(x,\t) &:= \frac{V(x,\t)-V(x,\t-\d\t)}{\d\t}
\end{align*} 
and $a,b,c,d$ are functions of $x$ and $\t$. Now demanding that
$V=x$ and $V=1$ are solutions of the scheme corresponds to setting
$b=0, c=0$. Next, demanding that a given tradable $R(x,\t)$ is also
a solution leads to
\[ 
    a \d_x^+ \d_x^- R(x,\t) = d \d_\t^- R(x,\t) 
\]
Therefore, the fully implicit scheme, implied by the tradables
is given by
\begin{equation} 
    \label{eq:imp} 
    \Theta^-(x,\t) \d_x^+ \d_x^- V(x,\t) = \Gamma(x,\t) \d_\t^- V(x,\t)
\end{equation}
where
\begin{align*}
    \Theta^-(x,\t) &:= \d_\t^- R(x,\t) \\
    \Gamma(x,\t)   &:= \d_x^+ \d_x^- R(x,\t)
    \end{align*}
A remarkably simple scheme, by virtue of the fact that in
our formulation of the pricing problem there are no drift terms
in the PDE. Of course we can also derive an explicit scheme along the same lines. 
We find
\begin{equation} 
    \label{eq:exp} 
    \Theta^+(x,\t) \d_x^+ \d_x^- V(x,\t) = \Gamma(x,\t) \d_\t^+ V(x,\t)
    \end{equation}
where now
\begin{align*}
    \d_\t^+ V(x,\t) &:= \frac{V(x,\t+\d\t)-V(x,\t)}{\d\t} \\
    \Theta^+(x,\t)  &:= \d_\t^+ R(x,\t)
\end{align*}
To further improve the accuracy of the scheme, we have combined the
two schemes into a Crank-Nicolson type of scheme, i.e. we take the
sum of Eqs.~\ref{eq:imp},\ref{eq:exp}, taking care of a proper time
shift in one of them.

It is not difficult to prove that our scheme is of order $\cO(\delta\tau^2)+\cO(\delta x^2)$, stable, and convergent. This is discussed in detail in appendix~\ref{sec:conv}.

Note that the scheme deviates from the usual Crank-Nicolson scheme in that the volatility function $\sigma(x,\tau)$ is approximated in the finite-difference scheme such that the tradable $R(x,\tau)$ is solved exactly
\begin{equation}
    \frac{\Theta^+(x,\tau)+\Theta^-(x,\tau)}{2\Gamma(x,\tau)} = \half\sigma(x,\tau)^2x^2 + \cO(\delta\tau^2)+\cO(\delta x^2)
\end{equation}
So the fitting of the scheme to the tradable can be understood as adding a correction to the volatility function that specifies the model. Note also that this scheme uses a different sort of fitting than what is normally done in a so-called fitted scheme in the literature.

\section{Application to a specific PDE}

A very interesting application of the scheme is in finding
solutions of the PDE
\begin{equation}
  \label{eq:pde}
  \bigg(
    -\partial_\t + \half \s^2 (x-\phi(\t))^2 \partial_x^2 
  \bigg)
  V(x,\t)=0
\end{equation}
where $\phi(\t)$ is a deterministic function. As was shown in our
articles Ref.~\cite{HooglandNeumann00b, HooglandNeumann00c}, this PDE
can be used to price arithmetic Asian options as well as plain vanilla
options on stocks paying cash dividends. Let us briefly recall how the
PDE is related to such options. To price an arithmetic average strike
put (unseasoned) with a payoff defined by
\begin{equation}
  \label{eq:asp}
  \left( \int_0^\t w(u)\frac{S(u)}{P(u)}du - kS \right)^+ 
\end{equation}
we set
\[ 
    \phi(\t) = \int_0^\t w(u) du 
\]
and then solve the PDE with boundary condition
\[ 
    V(x,0) = (x-k)^+ 
\]
The price then follows from
\[ 
    \mbox{Price} = S V(\phi(\t),\t) 
\]
The PDE can also be used to price average price options (using
time reversal duality) and seasoned Asians in general. Details
can be found in Ref.~\cite{HooglandNeumann00c}. To price a vanilla
option (say a call) with strike $K$ on a stock paying a stream
of cash dividends with density $w(\t)$, expressed in units of
the bond, we set
\[ 
    \phi(\t) = - \int_0^\t w(u) du 
\]
(notice the sign) and then solve the PDE with boundary condition
\[ 
    V(x,0) = (x-K)^+ 
\]
The price then follows from
\[ 
    \mbox{Price} = P V\big(S/P+\phi(\t),\t\big) 
\]
Note that if dividend payments are discrete (or for the Asian,
sampling is discrete), $w(\t)$ becomes a weighted sum of Dirac
delta-functions, and $\phi(\t)$ becomes a piece-wise constant
function, making jumps at dividend (sample) dates.

\vspace{1\baselineskip} \noindent Now the key question is: can we find
exact solutions to the PDE, simple enough, which can be used to set up
our implied mixed scheme?  Fortunately, the answer is yes. Again the
idea is to use a quadratic tradable, equalling $x^2$ at $\t=0$. This
is a good choice for two reasons. Firstly it is the simplest
non-trivial tradable. Secondly it has an $x$-independent gamma, which
is convenient for numerical reasons. Assuming that the solution
remains a quadratic polynomial in $x$ we try a solution of the form
\[ 
    R(x,\t) = A(\t)x^2+B(\t)x+C(\t) 
\]
Plugging this into Eq.~\ref{eq:pde}, we find the following
relations for the functions $A,B,C$:
\begin{align*}
    \partial_\t A(\t) &=    \s^2            A(\t) \\
    \partial_\t B(\t) &= -2 \s^2 \phi(\t)   A(\t) \\
    \partial_\t C(\t) &=    \s^2 \phi(\t)^2 A(\t)
\end{align*}
and they have the following solution
\begin{align*}    
    A(\t) &= e^{\s^2\t} \\
    B(\t) &= -2 \s^2 \int_0^\t \phi(u)e^{\s^2 u} du \\
    C(\t) &= \s^2 \int_0^\t \phi(u)^2 e^{\s^2 u} du
\end{align*}
These equations can easily be solved exactly, both in the case
of discrete sampling (where $\phi$ is piecewise constant) and
continuous sampling. For example, if we have a continuous
exponential weight function $w(\t)=e^{-r\t}/T$, we have
\[ 
\phi(\t)=\frac{1-e^{-r\t}}{rT} 
\]
and
\begin{align*}
    B(\t) &= \frac{2}{T(r-\s^2)}
    - \frac{2e^{\s^2\t}}{rT}\bigg( 1+\frac{e^{-r\t}\s^2}{r-\s^2} \bigg) \\
    C(\t) &= -\frac{2}{T^2(r-\s^2)(2r-\s^2)} 
    + \frac{e^{\s^2\t}}{r^2T^2}\bigg( 1+\frac{2e^{-r\t}\s^2}{r-\s^2}-\frac{e^{-2r\t}\s^2}{2r-\s^2}\bigg)
\end{align*}

\section{Results}

We have implemented the mixed scheme in C++. The numerical tests were
performed on a Pentium 233MHz machine. We have considered various
Asian options as well as options on stocks paying cash dividends. For
each option, we have compared results for three different grid sizes.
The results coming from the largest grid are so much more accurate
than the ones that we could find in the literature, that we have used
this as benchmark. We claim that these results are accurate to at
least six significant digits. The results compare well with numbers
recently obtained in Ref.~\cite{Zhang01}. The grid-sizes that we have
used are summarized in the table.

\begin{center} \begin{tabular}{|c|c|}
    \hline 
    Name & $N_x N_\t$ \\ \hline
    ``Exact'' & $10^8$ \\ \hline
    1 sec & $10^6$ \\ \hline
    10 ms & $10^4$ \\ \hline
    \end{tabular} \end{center}

\noindent The time labels of course indicate the execution speed
of the algorithm on our computer. The total number of points on
the grid were divided over the $x$ and $\t$ direction in such a
way that $N_\t/N_x\sim\s^2 T$. In all cases, care was taken that 
non-differentiable points on the payoff (at the strike) fell
on gridpoints. The boundary conditions in the $\t$-direction
were placed at zero and twice the strike. They are taken to be
time-independent (since gamma is very small there). So for a
payoff of the form
\[ 
    V(x,0) = (x-k)^+ 
\]
we also imposed
\[ 
    V(0,\t) = 0, \hspace{5mm} V(2k,\t)=k 
\]
Note that the scheme can be improved on this point: for
high volatility it might be necessary to place the bounds
further away from the strike, while for low volatility the
scheme can be improved by moving them closer. This leads to
a speedup of at least a factor 5 for the lowest volatilities.

\vspace{1\baselineskip} \noindent
The first instrument we consider is a standard
continuously sampled average strike Asian put option.
This corresponds to the choice of weight function
$w(\t)=e^{-r\t}/T$ (see Eq.~\ref{eq:asp} with $k=1$). We
use the parameters values used in Ref.~\cite{RogersShi95}:
initial stockprice $S=100$ and expiry time is 1 year.
The results in Table 1 should be compared to the results quoted in Table 2 of Ref.~\cite{Thompson99}.
Note that except for the lowest volatility setting, the
1 second results are practically equal to the exact
ones. As was mentioned before, the results can be
improved by using tighter boundary conditions for low
volatility.

\begin{center} 
\begin{tabular}{|c|c|c|c|c|c|}
\hline $\s$ & $r$ & Exact & 1 sec & 10 ms & Ref.\cite{Thompson99}\\ \hline
    & 0.05 & 1.24546 & 1.24546 & 1.24617 & 1.2454 \\ \cline{2-6}
0.1 & 0.09 & .699292 & .699311 & .701360 & 0.6992 \\ \cline{2-6}
    & 0.15 & .251676 & .251712 & .255256 & 0.2516 \\ \hline\hline
    & 0.05 & 3.40481 & 3.40481 & 3.40500 & 3.4044 \\ \cline{2-6}
0.2 & 0.09 & 2.62202 & 2.62202 & 2.62224 & 2.6216 \\ \cline{2-6}
    & 0.15 & 1.71019 & 1.71021 & 1.71193 & 1.7098 \\ \hline\hline
    & 0.05 & 5.62603 & 5.62602 & 5.62592 & 5.6246 \\ \cline{2-6}
0.3 & 0.09 & 4.73955 & 4.73954 & 4.73940 & 4.7382 \\ \cline{2-6}
    & 0.15 & 3.60981 & 3.60981 & 3.60985 & 3.6085 \\ \hline
\end{tabular} 
\\[1\baselineskip] 
Table 1: Continuously sampled average strike put
\end{center}

\noindent To see how the scheme performs in a discrete
sampling setting, we next consider a discretely sampled
average strike Asian put. Parameters settings are $S=100$,
$\s=0.2$, $r=0.1$ and expiry time is 1 year. There are $N$
sample dates, all with weight $1/N$ in terms of the money
value of the stock and taken at the beginning of each of
the $N$ equal length time periods in which the year is
divided. This corresponds to a weight function
\[ 
    w(\t) = \sum_{i=1}^N \frac{e^{-r\t}}{N} \d\left(\t-\frac{i}{N}\right) 
\]
Note that for $N=1$ the option is essentially a plain
vanilla put with strike $100$. For $N\to\infty$, the
option becomes a continuously sampled Asian. It must be noted
that in most occasions sample dates do not correspond to
grid-points in the time direction. In some cases there are many
more grid-points than there are samples, in other cases
the reverse is true. Obviously, the scheme handles both
cases very well. This is due to the fact that the scheme
is fitted to an exact solution of the PDE.

\begin{center} 
\begin{tabular}{|c|c|c|c|} \hline
$N$ & Exact & 1 sec & 10 ms \\ \hline
       1 & 3.75342 & 3.75342 & 3.75334 \\ \hline
       2 & 3.12047 & 3.12047 & 3.12062 \\ \hline
       4 & 2.79055 & 2.79055 & 2.79106 \\ \hline
       8 & 2.62151 & 2.62151 & 2.62217 \\ \hline
      16 & 2.53578 & 2.53578 & 2.53645 \\ \hline
      32 & 2.49257 & 2.49257 & 2.49324 \\ \hline
      64 & 2.47088 & 2.47088 & 2.47153 \\ \hline
     128 & 2.46001 & 2.46001 & 2.46065 \\ \hline
     256 & 2.45456 & 2.45457 & 2.45521 \\ \hline
     512 & 2.45184 & 2.45185 & 2.45249 \\ \hline
    1024 & 2.45048 & 2.45048 & 2.45112 \\ \hline
$\infty$ & 2.44912 & 2.44912 & 2.44976 \\ \hline
\end{tabular}
\\[1\baselineskip] 
Table 2: Discretely sampled average strike put
\end{center}

\noindent Since prices of average price Asians can be
related to prices of average strike Asians by T(ime-reversal)
duality \cite{HooglandNeumann00c}, it is not really neccessary
to give results for average price Asians. But in order to
allow the reader to compare our result to those in the
literature, we will give the results anyway. We consider
standard continuously sampled average price calls with parameter
settings $S=100$, $r=0.09$ and expiry in 1 year. The results in Table~\ref{tab:apc_cont} 
should be compared to the results quoted in Table 3 of Ref.\cite{RogersShi95} and Tables 5, 6, and 7 of Ref.\cite{Zhang01}.

\begin{center} 
\label{tab:apc_cont}
\begin{tabular}{|c|c|c|c|c|c|}
\hline $\s$ & $K$ & Exact & 1 sec & 10 ms & Ref.\cite{Zhang01} \\ \hline
     &  95 & 8.80884 & 8.80884 & 8.80895 & 8.80884 \\ \cline{2-6}
0.05 & 100 & 4.30823 & 4.30825 & 4.31013 & 4.30824 \\ \cline{2-6}
     & 105 & 0.958384 & 0.958385 & 0.958493 & 0.95838 \\ \hline\hline
     &  95 & 8.91185 & 8.91189 & 8.91538 & 8.91185 \\ \cline{2-6}
 0.1 & 100 & 4.91512 & 4.91515 & 4.91802 & 4.91512 \\ \cline{2-6}
     & 105 & 2.07006 & 2.07006 & 2.07022 & 2.07006 \\ \hline\hline
     &  90 & 14.9840 & 14.9841 & 14.9951 & 14.98397 \\ \cline{2-6}
 0.3 & 100 & 8.82876 & 8.82880 & 8.83275 & 8.82876 \\ \cline{2-6}
     & 110 & 4.69671 & 4.69670 & 4.69640 & 4.69670 \\ \hline\hline
     &  90 & 18.1886 & 18.1888 & 18.2055 &  \\ \cline{2-6}
 0.5 & 100 & 13.0281 & 13.0282 & 13.0357 &  \\ \cline{2-6}
     & 110 & 9.12429 & 9.12429 & 9.12454 &  \\ \hline
\end{tabular}
\\[1\baselineskip] 
Table 3: Continuously sampled average price call
\end{center}

\noindent As a final example we consider plain vanilla
options on a stock paying cash dividend. We consider a stock
with initial value $S=100$ and volatility $\s=0.1$. The
stock will pay a cash dividend of 5 after half a year.
Interest rate $r=0.05$. We consider a put and a call option
on this stock expiring in 1 year. In this case we
not only look at European style but also at American style
options. To calculate the latter we take the maximum
of the early exercise payoff and the calculated option
value at each time-step. For a call, the early exercise payoff
takes the form
\[ 
\big( x-\phi(\t)-Ke^{r\t} \big)^+ 
\]
in our coordinates. (Note that we can also use this technique
to valuate American style average strike Asians). The results
show that convergence is slower in the American case. This can
be explained by the fact that the early exercise feature is only
enforced at grid-points, in effect approximating the American
with a Bermudan option. Therefore the value is systematically
underestimated. Of course this is a general problem when
evaluating American options with finite difference schemes.
Nevertheless, the accuracy of our scheme is still of order $0.5\%$
for such options in the 10 ms grid-size.

\begin{center} 
\begin{tabular}{|c|c|c|c|c|c|c|} \hline
\multicolumn{1}{|c|}{Call} & 
\multicolumn{3}{c|}{European} &
\multicolumn{3}{c|}{American} \\ \hline
$K$ & Exact & 1 sec & 10 ms & Exact & 1 sec & 10 ms \\ \hline
 95 & 6.63807 & 6.63808 & 6.63910 & 8.00342 & 8.00405 & 7.97394 \\ \hline
100 & 3.89199 & 3.89199 & 3.89207 & 4.55708 & 4.55755 & 4.52858 \\ \hline
105 & 2.05220 & 2.05220 & 2.05248 & 2.28620 & 2.28646 & 2.27272 \\ \hline
\end{tabular} \end{center}

\begin{center} 
\begin{tabular}{|c|c|c|c|c|c|c|} \hline
\multicolumn{1}{|c|}{Put} & 
\multicolumn{3}{c|}{European} &
\multicolumn{3}{c|}{American} \\ \hline
$K$ & Exact & 1 sec & 10 ms & Exact & 1 sec & 10 ms \\ \hline
 95 & 1.88141 & 1.88142 & 1.88244 & 2.26310 & 2.26192 & 2.25389 \\ \hline
100 & 3.89148 & 3.89148 & 3.89156 & 4.72489 & 4.72226 & 4.70623 \\ \hline
105 & 6.80784 & 6.80784 & 6.80812 & 8.22660 & 8.22212 & 8.20011 \\ \hline
\end{tabular}
\\[1\baselineskip] 
Table 4: Vanilla options with cash dividends
\end{center}

\section{Conclusions and outlook}

In this article we introduced a very simple and natural
finite-difference scheme for pricing derivative securities.  The
scheme is easy to implement and very flexible.  No special tricks need
to be used to reach a high accuracy.  The scheme follows in a
straightforward manner from our formulation of the pricing problem in
terms of tradables. It is (at least) accurate to order $\cO(\delta\tau^2)+\cO(\delta x^2)$, stable, and convergent.

The key idea is that one can often find {\it
  exact} solutions to the governing PDE. These solutions can be used
to tune the finite-difference scheme in a very simple manner.
Problems usually associated to the presence of drift terms, e.g.
boundary layers, are trivially avoided. Numerical computations for
arithmetic Asian options show that convergence is fast and neither the
solution nor the greeks are plagued by oscillations. Also, financial
instruments often depend on values of underlying on discrete points in
time, and this can lead to discretization errors in a finite
difference scheme when these points do not match with gridpoints. Our
algorithm compensates for this problem by making use of exact
solutions.

Although this article focussed on adapting a finite-difference
scheme to exact solutions of a particular pricing PDE, it does not take
too much effort to extent our method in such a way that observed
prices are used to construct an implied finite difference scheme,
extending the idea of implied trees. We will come back to this in a
future article.

\section*{Acknowledgements}

We would like to thank Daniel Duffy and Piet Hemker for extensive
discussions on boundary layer problems, fitting schemes, and singular
perturbation methods.

\appendix
\section{Convergence of the scheme}
\label{sec:conv}
In this appendix we give a proof of convergence of the numerical scheme proposed in this article, under the following conditions on the tradable $R(x,\tau)$ used to fix the scheme
\begin{subequations}
\label{eq:cond}
\begin{align}
    \partial_\tau R(x,\tau) &\ge 0 \\
    \partial_x^2 R(x,\tau) &> 0
\end{align}
\end{subequations}
for all $x_{\min}\le x\le x_{\max}$ and $\tau_{\min}\le \tau\le \tau_{\max}$. In other words, $R(x,\tau)$ should be a strictly convex security. We will assume that the solution $V(x,\tau)$ has vanishing gamma (second derivative w.r.t. $x$) at the boundaries $x=x_{\min}$ and $x=x_{\max}$. Such boundary conditions were used in all calculations in this article.
Later, we will briefly consider more general boundary conditions, which are needed, for example, in the pricing of lookback options. Now first we note that our scheme is consistent, i.e. in the limit $\delta\tau, \delta x\to0$, the defining equations reduce to the actual PDE that we want to solve. Then, by a theorem of Lax, the scheme will be convergent if it is stable, i.e. all eigenvalues of the matrix $M$, defining the vector $v$ of the values at time $\tau$ in terms of those at time $\tau-\delta\tau$
\begin{equation}
    \label{eq:mat}
    v(\tau) = M v(\tau-\delta\tau)
\end{equation}
have absolute value less than or equal to one. We now switch to index notation
\begin{align*}
    r_{i,j} &:= R(x_i,\tau_j) \\    
    v_{i,j} &:= v(x_i,\tau_j) \\    
    x_i &:= x_{\min} + i\frac{x_{\max}-x_{\min}}{N_x} \\
    \tau_j &:= \tau_{\min} + j\frac{\tau_{\max}-\tau_{\min}}{N_\tau}
    \end{align*}
where $i=0,\ldots,N_x$ and $j=0,\ldots,N_{\tau}$. For $i=1,\ldots,N_x-1$, the implicit equation becomes
\begin{align*}
\MoveEqLeft    
    \big(r_{i,j}-r_{i,j-1}\big)\big(v_{i-1,j}-2v_{i,j}+v_{i+1,j}\big) \\
    &= \big(v_{i,j}-v_{i,j-1}\big)\big(r_{i-1,j}-2r_{i,j}+r_{i+1,j}\big)
\end{align*}
or
\begin{align*}
    &\delta_\tau^- r_{i,j}\, \delta_x^+\delta_x^- v_{i,j} = \delta_\tau^- v_{i,j}\, \delta_x^+\delta_x^- r_{i,j}
\end{align*}
where we defined
\begin{align*}
    \delta_x^\pm v_{i,j} &:= v_{i\pm 1,j}-v_{i,j} \\
    \delta_\tau^- v_{i,j} &:= v_{i,j}-v_{i,j-1}
\end{align*}
The explicit equation (after a time shift) becomes
\begin{align*}
\MoveEqLeft    
    \big(r_{i,j}-r_{i,j-1}\big)\big(v_{i-1,j-1}-2v_{i,j-1}+v_{i+1,j-1}\big) \\
    &= \big(v_{i,j}-v_{i,j-1}\big)\big(r_{i-1,j-1}-2r_{i,j-1}+r_{i+1,j-1}\big)
\end{align*}
or
\begin{align*}
    &\delta_\tau^- r_{i,j}\, \delta_x^+\delta_x^- v_{i,j-1} = \delta_\tau^- v_{i,j}\, \delta_x^+\delta_x^- r_{i,j-1}
\end{align*}
The Crank-Nicolson scheme is their sum. After rearranging
\begin{align*}
\MoveEqLeft    
    -\theta_{i,j} v_{i-1,j} + \big(\g_{i,j}+2\theta_{i,j}\big)v_{i,j} - \theta_{i,j}v_{i+1,j} \\
    &= \theta_{i,j} v_{i-1,j-1} + \big(\g_{i,j}-2\theta_{i,j}\big)v_{i,j-1} + \theta_{i,j}v_{i+1,j-1}
\end{align*}
where we introduced
\begin{eqnarray*}
    \gamma_{i,j} &:=& \big(r_{i-1,j}-2r_{i,j}+r_{i+1,j}\big)+\big(r_{i-1,j-1}-2r_{i,j-1}+r_{i+1,j-1}\big) \\
    \theta_{i,j} &:=& r_{i,j}-r_{i,j-1}
\end{eqnarray*}
The vanishing of gamma at the boundaries is implemented by demanding that 
$v_{i-1,j}-2v_{i,j}+v_{i+1,j}=0$ for $i=0$ and $i=N_x$. Inserting this in the scheme, this translates into
\begin{subequations} 
    \label{eq:bc}
\begin{align}
    v_{0,j} &= v_{0,j-1} \\
    v_{N_x,j} &= v_{N_x,j-1} 
\end{align}
\end{subequations}
In matrix notation, the entire scheme now becomes:
\[
    \big(\Gamma^j+\Theta^j\big)v_j = \big(\Gamma^j-\Theta^j\big)v_{j-1}
\]
where $v_j=\{v_{0,j},\ldots,v_{N_x,j}\}$, the diagonal matrix $\Gamma^j$ is defined by
\[
    \Gamma^j_{h,i} :=\gamma_{h,j}\delta_{h,i}
\]
and the tridiagonal matrix $\Theta^j$ is defined by
\[
    \Theta^j_{h,i} =
    \begin{cases}
    \theta_{h,j}\big(-\delta_{h,i+1}+2\delta_{h,i}-\delta_{h,i-1}\big) & \text{if } i=1,\ldots,N_x-1 \\
    0 & \text{if } i=0 \text{ or } i=N_x
    \end{cases}
\]
This neatly incorporates the boundary conditions (\ref{eq:cond}). By the assumptions Eq.~(\ref{eq:bc}) we have
\begin{subequations}    
    \label{eq:elem}
\begin{align}
    \gamma_{i,j} &= \int_{x_i}^{x_i+\delta x} dx' \int_{x'-\delta x}^{x'} dx \big(\partial_x^2 R(x,\tau_j) + \partial_x^2 R(x,\tau_j-\delta\tau)\big) > 0 \\
    \theta_{i,j} &= \int_{\tau_j-\delta\tau}^{\tau_j} d\tau \partial_\tau R(x_i,\tau) \ge 0
\end{align}
\end{subequations}
and this shows that $\Gamma^j$ is non-degenerate, so we can multiply both sides by the inverse of this matrix to get
\[
    \big({\bf 1}+(\Gamma^j)^{-1}\Theta^j\big) v_j =
    \big({\bf 1}-(\Gamma^j)^{-1}\Theta^j\big) v_{j-1}
\]
So the matrix which defines a timestep in Eq.~\ref{eq:mat} is given by
\[
    M := \big({\bf 1} + Q\big)^{-1}\big({\bf 1} - Q\big),
    \quad Q :=(\Gamma^j)^{-1}\Theta^j
\]
Consider an eigenvector of the matrix $Q$ with eigenvalue $\lambda$. Obviously, this will also be an eigenvector of $M$ with eigenvalue 
\[
    \frac{1-\lambda}{1+\lambda}
\]
The absolute value of this eigenvalue will be less than or equal to one iff $\lambda\ge0$. So we have to prove that $Q$ is positive semi-definite. As a matter of fact we can write
\[
    Q = 
    \begin{pmatrix}
         0 & 0 & \hdots & 0 & 0 \\
         0 & \frac{\theta_{1,j}}{\gamma_{1,j}} &  & 0 & 0 \\
    \vdots &   & \ddots &  & \vdots \\
         0 & 0 &  & \frac{\theta_{N_x-1,j}}{\gamma_{N_x-1,j}} & 0 \\
         0 & 0 & \hdots & 0 & 0 
    \end{pmatrix}
    \begin{pmatrix}
     2 & -1 &  0 &  0 & \hdots \\
    -1 &  2 & -1 &  0 &  \\
     0 & -1 &  2 & -1 &  \\
     0 &  0 & -1 &  2 &  \\
    \vdots &   &   &     & \ddots 
    \end{pmatrix}
\]
The first matrix is positive semi-definite by Eqs.~\ref{eq:elem}. The second is well known to be positive definite. Consequently, $Q$ is positive semi-definite and this concludes the proof of convergence given our assumptions. 

Note that for Asians contracts, we used a quadratic tradable $R$ which can be written as
\[
    R(x,\tau) = x^2+\sigma^2 \int_0^\tau \big(x-\phi(u)\big)^2e^{\sigma^2 u}du
\]
Therefore
\begin{align*}
    \partial_\tau R(x,\tau) &= \sigma^2 e^{\sigma^2\tau}\big(x-\phi(\tau)\big)^2 \ge 0\\
    \partial_x^2 R(x,\tau) &= 2 e^{\sigma^2\tau} > 0
\end{align*}
and the scheme is convergent for this particular case.

We conclude with a brief remark about a more general type of boundary conditions. Suppose we demand
\[
\big(a\partial_x^2+b\partial_x+c\big)V(x,\tau) = 0
\]
for $x=x_{\min}$. In terms of finite differences, this becomes
\[
\frac{a}{\delta x^2} \big(v_{1,j}-2v_{0,j}+v_{-1,j}\big)
+ \frac{b}{2\delta x} \big(v_{1,j}-v_{-1,j}\big)
+ c v_{0,j}
= 0
\]
Solving this for $v_{-1,j}$ we get (assuming $2a\ne b\delta x$)
\[
    v_{-1,j} = \mu v_{0,j} - \nu v_{1,j}, \quad
    \mu := \frac{4a-2c\delta x^2}{2a-b\delta x}, \quad
    \nu := \frac{2a+b\delta x}{2a-b\delta x}.
\]
This leads to the following condition in the CN-scheme
\begin{align*}
\MoveEqLeft
    \big(\gamma_{0,j}+(2-\mu)\theta_{0,j}\big)v_{0,j}-(1-\nu)\theta_{0,j}v_{1,j} \\
    &= \big(\gamma_{0,j}-(2-\mu)\theta_{0,j}\big)v_{0,j-1}+(1-\nu)\theta_{0,j}v_{1,j-1}
\end{align*}
If we now repeat the steps of the previous proof, we end up with a matrix $Q$ given by
\[
    Q = 
    \begin{pmatrix}
         \frac{\theta_{0,j}}{\gamma_{0,j}} & 0 & \hdots & 0 & 0 \\
         0 & \frac{\theta_{1,j}}{\gamma_{1,j}} &  & 0 & 0 \\
    \vdots &   & \ddots &  & \vdots \\
         0 & 0 &  & \frac{\theta_{N_x-1,j}}{\gamma_{N_x-1,j}} & 0 \\
         0 & 0 & \hdots & 0 & 0 
    \end{pmatrix}
    \begin{pmatrix}
     2-\mu & -1+\nu &  0 &  0 & \hdots \\
    -1 &  2 & -1 &  0 &  \\
     0 & -1 &  2 & -1 &  \\
     0 &  0 & -1 &  2 &  \\
    \vdots &   &   &     & \ddots 
    \end{pmatrix}    
\]
This matrix is positive semi-definite iff $\mu-\nu \le 1$, and in this case, the scheme is convergent. This is true in particular for the case
\[
    \partial_x V(x,\tau) =\alpha V(x,\tau)
\]
when $\alpha \ge 0$, which is needed for lookback options.

\end{document}